\documentclass{article}
\usepackage{amssymb}
\usepackage{amsmath}
\usepackage{epsf}
\usepackage{epsfig}
\usepackage{floatflt}
\usepackage{multimedia}
\usepackage{color}

\input{tcilatex}
\begin{document}

\begin{center}

\bigskip

{\Huge A new solvable many-body problem of goldfish type}\bigskip

$^{\ast }$\textbf{Oksana Bihun}$^{1}$ and $^{+\lozenge }$\textbf{Francesco
Calogero}$^{2}\bigskip $

$^{\ast }$Department of Mathematics, University of Colorado, Colorado
Springs, USA

$^{+}$Physics Department, University of Rome \textquotedblleft La Sapienza''

$^{\lozenge }$Istituto Nazionale di Fisica Nucleare, Sezione di Roma

$^{1}$obihun@cord.edu

$^{2}$francesco.calogero@roma1.infn.it, francesco.calogero@uniroma1.it

\bigskip

\textit{Abstract}
\end{center}

A new solvable many-body problem of goldfish type is introduced and the
behavior of its solutions is tersely discussed. 


\section{Introduction}

\textbf{Notation 1.1}. Hereafter $N$ is (unless otherwise
indicated) an \textit{arbitrary integer}, $N\geq 2$, the (generally \textit{%
complex}) numbers $z_{n}\equiv z_{n}\left( t\right) $ are the dependent
variables, $t$ (``time'') is the independent variable, superimposed dots
denote time-differentiations, and indices such as $n,$ $m,$ $\ell $ run over
the integers from $1$ to $N$ unless otherwise indicated (see for instance
below in (\ref{IsoGold}) the limitation $\ell \neq n$ on the $N$ values of $%
\ell $). Below we often omit to indicate explicitly the $t$-dependence of
various quantities, when this can be done without causing misunderstandings.
Hereafter $N\times N$ matrices are denoted by upper-case \textbf{boldface}
letters (so that, for instance, the matrix $\mathbf{C}$ has the $N^{2}$
elements $C_{nm}$). Lower-case \textbf{boldface} letters stand for $N$-vectors (so
that, for instance, the $N$-vector $\mathbf{z}$ has the $N$ components $%
z_{n})$; and  the \textit{imaginary unit}  is denoted by $\mathbf{i}$ (so that $\mathbf{i}%
^{2}=-1$, and $\mathbf{i}$ is of course \textit{not} a $N$-vector!). We
occasionally use the Kronecker symbol, with its standard definition: $\delta
_{mn}=1$ for $m=n,$ $\delta _{mn}=0$ for $m\neq n$. And let us mention the
standard convention according to which an empty sum vanishes and an empty
product equals unity, i. e. $\sum_{j=J}^{K}\left( f_{j}\right) =0,$ $%
\prod\nolimits_{j=J}^{K}\left( f_{j}\right) =1\ $if\ $K<J$. $\blacksquare $

A prototypical ``goldfish'' many-body model \cite{C2001a} \cite{C1978} \cite%
{C2001} \cite{C2008} is characterized by the \textit{translation-invariant}
equations of motion 
\begin{subequations}
\begin{equation}
\ddot{z}_{n}=\mathbf{i}~\omega ~\dot{z}_{n}+\sum_{\ell =1,~\ell \neq
n}^{N}\left( \frac{2~\dot{z}_{n}~\dot{z}_{\ell }}{z_{n}-z_{\ell }}\right) ~.
\label{IsoGold}
\end{equation}%
A Hamiltonian producing these equations of motion reads as follows: 
\begin{equation}
H\left( \mathbf{\zeta };\mathbf{z}\right) =\sum_{n=1}^{N}\left[ \mathbf{i}%
~\omega ~z_{n}+\exp \left( \zeta _{n}\right) ~\prod\limits_{\ell =1,~\ell
\neq n}^{N}\left( z_{n}-z_{\ell }\right) ^{-1}\right] ~,  \label{HamIsoGold}
\end{equation}%
where of course the $N$ coordinates $\zeta _{n}\equiv \zeta _{n}\left(
t\right) $ are the canonical momenta corresponding to the canonical particle
coordinates $z_{n}\equiv z_{n}\left( t\right) $. The solution of the
corresponding initial-values problem is provided by the $N$ roots $%
z_{n}\equiv z_{n}\left( t\right) $ of the following, rather neat, \textit{%
algebraic} equation in the variable $z$: 
\begin{equation}
\sum_{\ell =1,~\ell \neq n}^{N}\left[ \frac{\dot{z}_{\ell }\left( 0\right) +%
\mathbf{i}~\omega ~z_{\ell }\left( 0\right) }{z-z_{\ell }\left( 0\right) }%
\right] =\frac{\mathbf{i}~\omega }{\exp \left( \mathbf{i}~\omega ~t\right) -1%
}~.  \label{SolIsoGold}
\end{equation}%
(Note that this is actually a \textit{polynomial} equation of degree $N$ in $%
z$, as seen by multiplying it by $\tprod\nolimits_{m=1}^{N}\left[
z-z_{m}\left( 0\right) \right] $). Hence this model is~\textit{isochronous} (whenever the
parameter $\omega$ is \textit{positive}, as we generally assume hereafter; the special
case $\omega=0$ is ``the'' prototypical, \textit{nonisochronous}, case...): all its
solutions are \textit{completely periodic}, with the period $T=2\pi/\omega$ or, possibly,
due to an exchange of the particle positions through the motion, with a period that is a 
(generally small, see \cite{GS2005}) \textit{integer multiple} of $T$.

Several \textit{solvable} generalizations of the goldfish model,
characterized by Newtonian equations of motion featuring additional forces
besides those appearing in the right-hand side of (\ref{IsoGold}), are
known: see for instance \cite{C1978} \cite{C2001} \cite{C2008} and
references therein.

\textbf{Remark 1.1}. Above and hereafter we call a many-body model \textit{%
solvable} if its initial-values problem can be solved by \textit{algebraic}
operations, such as finding the $N$ zeros of a known $t$-dependent \textit{%
polynomial} of degree $N$ (of course such an algebraic equation can be 
\textit{explicitly} solved only for $N\leq 4$). $\blacksquare $

Recently a simple technique has been introduced \cite{C2015}, which allows
to identify and investigate additional \textit{solvable} models of goldfish
type; and a few examples of such models yielded by this new approach have
been identified\ and tersely discussed \cite{C2015}. The model treated in
this paper is another such new model, which is perhaps itself \textit{%
interesting} (as all solvable models tend to be), and moreover allows---as
reported in a separate paper \cite{BC2015}---to obtain remarkable \textit{%
Diophantine} results for the \textit{zeros} of (monic) polynomials of degree 
$N$ the \textit{coefficients} of which are the \textit{zeros} of Hermite
polynomials (see, for instance, \cite{E1953}) of degree $N$.

\bigskip

\section{The model and its solutions}

The Newtonian equations of motion of the new many-body problem of goldfish
type read as follows: 
\end{subequations}
\begin{subequations}
\label{NewGold}
\begin{eqnarray}
&&\ddot{z}_{n}=\sum_{\ell =1,~\ell \neq n}^{N}\left( \frac{2~\dot{z}_{n}~%
\dot{z}_{\ell }}{z_{n}-z_{\ell }}\right) -\left[ \tprod\limits_{\ell
=1,~\ell \neq n}^{N}\left( z_{n}-z_{\ell }\right) \right] ^{-1}\cdot  \notag
\\
&&\cdot \sum_{m=1}^{N}\left\{ \left( z_{n}\right) ^{N-m}~\left[ -\omega
^{2}~c_{m}+2~\sum_{\ell =1,~\ell \neq m}^{N}\left( c_{m}-c_{\ell }\right)
^{-3}\right] \right\} ~,  \label{zndotdot}
\end{eqnarray}%
with%
\begin{equation}
c_{m}=\left( -1\right) ^{m}~\sum_{1\leq s_{1}<s_{2}...<s_{m}\leq N}\left[
\tprod\limits_{r=1}^{m}\left( z_{s_{r}}\right) \right] ~.  \label{cm}
\end{equation}%
Here and hereafter the symbol $\sum_{1\leq s_{1}<s_{2}...<s_{m}\leq N}$
signifies the sum from $1$ to $N$ over the $m$ (integer) indices $s_{j}$
with $j=1,...,m$ and the restriction $s_1<s_2<\ldots <s_m;$ of course this sum
vanishes if $m>N,$ consistently with \textbf{Notation 1.1}.

The solutions $z_{n}\equiv z_{n}\left( t\right) $ of this $N$-body problem
are provided---consistently with the expressions (\ref{cm})---by the $N$
zeros of the following $t$-dependent (monic) polynomial $\psi _{N}\left(
z;t\right) $ of degree $N$ in $z$: 
\end{subequations}
\begin{equation}
\psi _{N}\left( z;t\right) =z^{N}+\sum_{m=1}^{N}\left[ c_{m}\left( t\right)
~z^{N-m}\right] ~,  \label{psi}
\end{equation}%
where the coefficients $c_{m}\left( t\right) $ are themselves the solutions
of the system of $N$ ODEs%
\begin{equation}
\ddot{c}_{m}=-\omega ^{2}~c_{m}+2~\sum_{\ell =1,~\ell \neq m}^{N}\left(
c_{m}-c_{\ell }\right) ^{-3}~.  \label{cmdotdot}
\end{equation}%
Because this is a well-known \textit{solvable }model, the time-dependence of
these $N$ quantities $c_{m}\left( t\right) $ can be obtained by solving an 
\textit{algebraic} (in fact \textit{polynomial}) problem, indeed the
solution of the initial-value problem of this dynamical system, (\ref%
{cmdotdot}), is provided by the following prescription (see for instance section 4.2.2 in \cite{C2008} or
\cite{OP1981, C2001}): the $N$ quantities $c_{m}\equiv
c_{m}\left( t\right) $ are the $N$ \textit{eigenvalues} of the $N\times N$ ($%
t$-dependent) matrix 
\begin{subequations}
\begin{equation}
\mathbf{C}\left( t\right) =\mathbf{C}\left( 0\right) ~\cos (\omega ~t)+%
\mathbf{\dot{C}}\left( 0\right) ~\frac{\sin (\omega ~t)}{\omega }~,
\end{equation}%
with 
\begin{equation}
\mathbf{C}\left( 0\right) =\text{diag}\left[ c_{m}\left( 0\right) \right] ~,
\end{equation}%
\begin{equation}
\mathbf{\dot{C}}\left( 0\right) =\text{diag}\left[ \dot{c}_{m}\left(
0\right) \right] +\mathbf{i}~\left[ \mathbf{M}\left( 0\right) ,~\mathbf{C}%
\left( 0\right) \right] ~,  \label{Cdot(0)}
\end{equation}%
where of course (see (\ref{cm}))%
\begin{equation}
c_{m}\left( 0\right) =\left( -1\right) ^{m}~\sum_{1\leq
s_{1}<s_{2}...<s_{m}\leq N}\left\{ \tprod\limits_{r=1}^{m}\left[
z_{s_{r}}\left( 0\right) \right] \right\} ~,
\end{equation}%
\begin{equation}
\dot{c}_{m}\left( 0\right) =\left( -1\right) ^{m}~\sum_{1\leq
s_{1}<s_{2}...<s_{m}\leq N}~\sum_{q=1}^{m} \left\{ \dot{z}_{s_q}\left(
0\right) \tprod\limits_{r=1,~r\neq q}^{m}\left[ z_{s_{r}}\left( 0\right) %
\right] \right\}  ~,
\end{equation}%
and in the right-hand side of (\ref{Cdot(0)}) 
\begin{equation}
\left[ \mathbf{M}\left( 0\right) ,~\mathbf{C}\left( 0\right) \right] \equiv 
\mathbf{M}\left( 0\right) ~\mathbf{C}\left( 0\right) -\mathbf{C}\left(
0\right) ~\mathbf{M}\left( 0\right)
\end{equation}%
with the matrix $\mathbf{M}\left( 0\right) $ defined componentwise in terms
of the initial data $z_{n}\left( 0\right) $ as follows:%
\begin{eqnarray}
M_{nm}\left( 0\right) &=&\left[ z_{n}\left( 0\right) -z_{m}\left( 0\right) %
\right] ^{-2}~,~~~n\neq m~,  \notag \\
M_{nn}\left( 0\right) &=&-\sum_{\ell =1,~\ell \neq n}M_{n\ell }\left(
0\right) =-\sum_{\ell =1,~\ell \neq n}\left[ z_{n}\left( 0\right) -z_{\ell
}\left( 0\right) \right] ^{-2}~.
\end{eqnarray}

Note that these formulas provide an explicit definition of the $N$
time-dependent coefficients $c_{m}\left( t\right) $ in terms of the initial
data $z_{n}\left( 0\right) ,$ $\dot{z}_{n}\left( 0\right) $ of the $N$-body
problem of goldfish type characterized by the Newtonian equations of motion (%
\ref{NewGold}), via algebraic operations, amounting essentially to the
solution of polynomial equations of degree $N$; and that the values $%
z_{n}\left( t\right) $ at time $t$ of the particle coordinates $z_{n}$ are
then provided by the $N$ zeros of the polynomial $\psi _{N}\left( z;t\right) 
$, explicitly known (see (\ref{psi})) in terms of its $N$ coefficients $%
c_{m}\left( t\right) $. It is thereby demonstrated that the $N$-body problem
of goldfish type characterized by the Newtonian equations of motion (\ref%
{NewGold}) is \textit{solvable }(see \textbf{Remark 1.1}).

\textbf{Remark 2.1}. Let us call attention to a (well known) tricky point
associated with the solution---as described above---of the $N$-body problem
of goldfish type characterized by the Newtonian equations of motion (\ref%
{NewGold}). The identification of the $N$ eigenvalues of a given matrix is
only \textit{unique up to permutations}, and likewise the identification of
the zeros of a polynomial is only \textit{unique up to permutations}.
Therefore the $N$ coordinates $z_{n}=z_{n}\left( t\right) $ yielded by the
solution detailed above are only identified \textit{up to permutations of
their }$N$\textit{\ labels }$n$. The (only) way to identify a \textit{%
specific} coordinate---say, the coordinate $z_{1}\left( t\right) $ that
corresponds to the initial data $z_{1}\left( 0\right) ,$ $\dot{z}_{1}\left(
0\right) $---is by following the (\textit{continuous}) time evolution of the
coordinate $z_{1}\left( t\right) $ from its (assigned) initial value $%
z_{1}\left( 0\right) $ to its value $z_{1}\left( t\right) $ at time $t$. In
this manner one arrives at the \textit{uniquely }defined value of the
coordinate $z_{1}\left( t\right) $ corresponding to the initial value $%
z_{1}\left( 0\right) $, which coincides of course with that \textit{uniquely}
yielded by the time evolution of the $N$-body problem (\ref{NewGold}). 
An analogous phenomenology is also relevant to the solution of model~(\ref{IsoGold}).
$%
\blacksquare $

Because of the way system~(\ref{NewGold}) is 
constructed, its equilibria  can be obtained by finding the \textit{zeros} of the polynomials whose \textit{coefficients} are equilibria of system~(\ref{cmdotdot}). On the other hand, it is known that the \textit{zeros} of the $N$-th degree Hermite polynomial are equilibria of system~(\ref{cmdotdot}) (see for instance \cite{C2001}). This relationship allows to prove \textit{Diophantine} properties of the polynomials whose \textit{coefficients} are the \textit{zeros} 
of Hermite polynomials. These properties, together with the study of  the behavior of
system~(\ref{NewGold}) in the immediate vicinity of its equilibria, are discussed in a
separate paper \cite{BC2015}.

The findings reported above imply the possibility to display in \textit{%
completely explicit} form the solution of the $N$-body problem of goldfish
type characterized by the Newtonian equations of motion (\ref{NewGold}) for $%
N=2,$ $3,$ $4$; but we doubt this would be very illuminating, and we
therefore leave this task to the eager reader. We rather like to emphasize
that these findings imply that, for arbitrary $N$---and arbitrary positive $%
\omega $---this $N$-body problem is \textit{isochronous}, all its solutions
satisfying the periodicity property 
\end{subequations}
\begin{equation}
z_{n}\left( t+T\right) =z_{n}\left( t\right) ~,
\end{equation}%
with $T=2\pi /\omega $ (or possibly $T$ might be replaced by a, generally
small, integer multiple of $2\pi /\omega $: see \cite{GS2005}). We end this paper by
displaying a few examples of this phenomenology. 

\bigskip

\textbf{Example 2.1}.  For $N=2$ and $\omega=1$, taking into account that $c_1=-z_1-z_2$ and $c_2=z_1 z_2$, we see that system~(\ref{NewGold}) reduces to
\begin{eqnarray}
&&\ddot{z}_1=\frac{2 \dot{z}_1 \dot{z}_2}{z_1-z_2}-\frac{1}{(z_1-z_2)}\left[
 z_1^2-\frac{2(z_1-1)}{(z_1+z_2+z_1 z_2)^3}
 \right] , \notag\\
 &&\ddot{z}_2=-\frac{2 \dot{z}_1 \dot{z}_2}{z_1-z_2}+\frac{1}{(z_1-z_2)}\left[
 z_2^2-\frac{2(z_2-1)}{(z_1+z_2+z_1 z_2)^3}
 \right] .\notag\\
\label{znsystN2}
\end{eqnarray}
The assignment $\omega=1$ is motivated by the fact that this quantity, $\omega$, can be eliminated from system~(\ref{IsoGold})
by a \textit{constant} rescaling of the dependent variables $z_n(t)$ and of the independent variable $t$.

System~(\ref{znsystN2}) has 4 equilibrium configurations $(\hat{z}^{(j)}_1, \hat{z}^{(j)}_2)$, $j=1,2,3,4$, up to the exchange of $\hat{z}^{(j)}_1$ with $\hat{z}^{(j)}_2$, whose approximate numerical values are given below:
\begin{eqnarray}
\begin{array}{l}
(\hat{z}^{(1)}_{1}, \hat{z}^{(1)}_{2})=(0.353553 - 0.762959~\mathbf{i}, 0.353553 + 0.762959~\mathbf{i}),\\
(\hat{z}^{(2)}_{1}, \hat{z}^{(2)}_{2})=(-0.54455 + 1.00281~\mathbf{i}, 0.54455 - 0.295704~\mathbf{i}),\\
(\hat{z}^{(3)}_{1}, \hat{z}^{(3)}_{2})=(-0.54455 - 1.00281~\mathbf{i}, 0.54455 + 0.295704~\mathbf{i}),\\
(\hat{z}^{(4)}_{1}, \hat{z}^{(4)}_{2})=(-1.26575, 0.558645).
\end{array}
\label{EquiSystznN2}
\end{eqnarray}
These equilibria can be obtained either by  the substitution $z_n(t)=\hat{z}_n$, $n=1,2$ into system~~(\ref{znsystN2}) and the subsequent solution of the resulting system of algebraic equations for $\hat{z}_1,\hat{z}_2$, or by finding the zeros of the polynomials whose coefficients are the equilibria of system~(\ref{cmdotdot}) for $N=2$. We note that, in this case where $N=2$ and $\omega=1$, system (\ref{cmdotdot}) has the four equilibria $(\frac{1}{\sqrt{2}},-\frac{1}{\sqrt{2}})$, $(-\frac{1}{\sqrt{2}},\frac{1}{\sqrt{2}})$,
$(\frac{\mathbf{i}}{\sqrt{2}},-\frac{\mathbf{i}}{\sqrt{2}})$, and $(-\frac{\mathbf{i}}{\sqrt{2}},\frac{\mathbf{i}}{\sqrt{2}})$. Two of them are the zeros $\pm\frac{1}{\sqrt{2}}$ of the Hermite polynomial $H_2(c)=4 c^2-2$, which is consistent with the known fact that the zeros of the $N$-th order Hermite polynomial are the equilibria of system~(\ref{cmdotdot}).

\begin{figure}
\includegraphics[width=7cm]{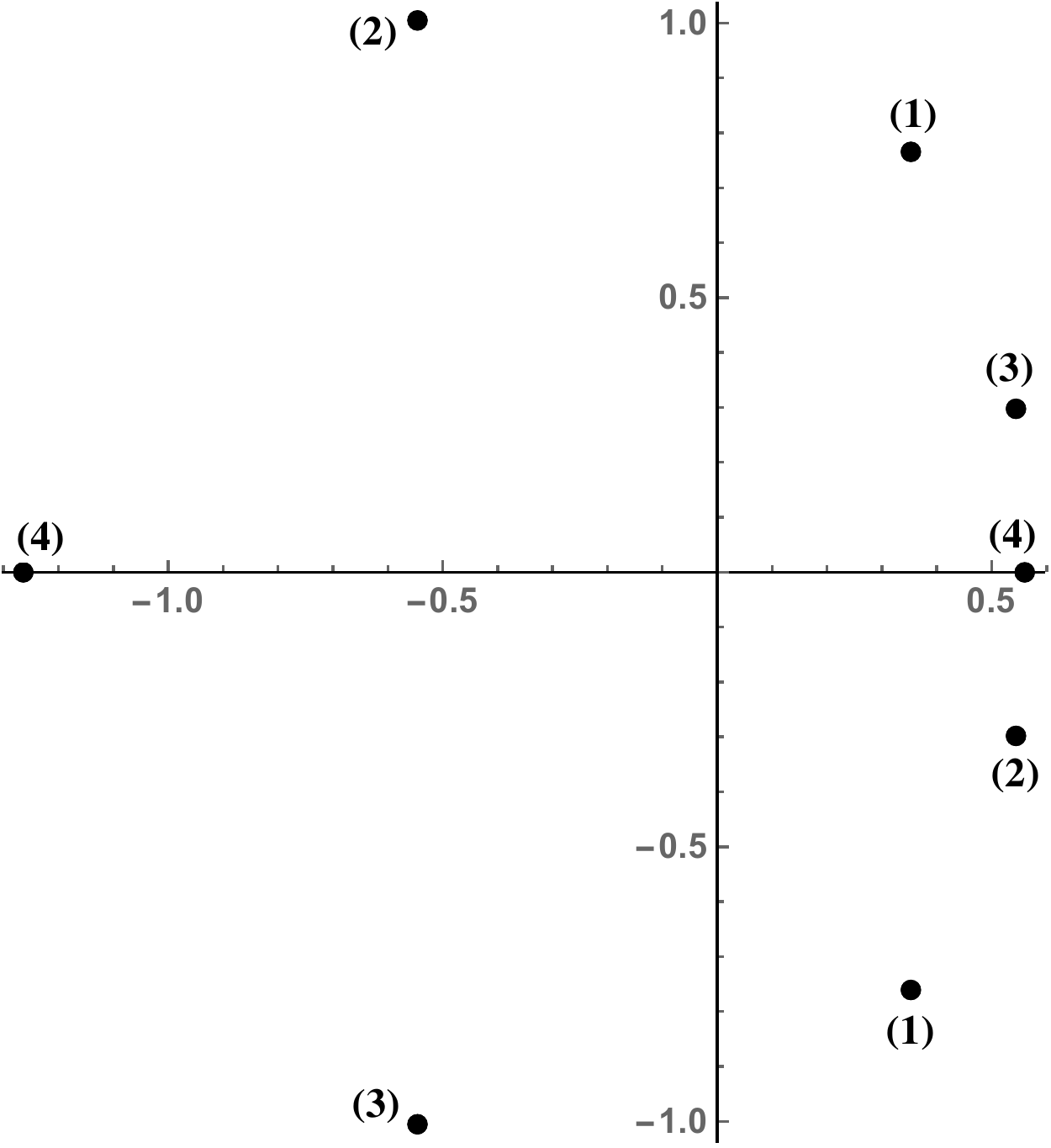}
\caption{Equilibria of system~(\ref{znsystN2}). Each equilibrium is represented by the two points $\hat{z}^{(j)}_{1}$ and $\hat{z}^{(j)}_{2}$ labeled by the index $j=1,2,3,4$, see~(\ref{EquiSystznN2}).}
\label{Equilibria}
\end{figure}

In Figure~\ref{Equilibria} each equilibrium of system~(\ref{znsystN2}) is represented by the two points $\hat{z}^{(j)}_{1}$ and $\hat{z}^{(j)}_{2}$, labeled by $(j)$, where $j=1,2,3,4$.

Below we provide  
\textit{graphs} of the real and imaginary parts of the components of the solution $%
(z_{1}\left( t\right) , z_2(t))$ of system~(\ref{znsystN2}) as functions of time, and some \textit{trajectories}
of the particles $z_1$ and $z_2$ in the complex $z$-plane. These graphs have been obtained by solving system~(\ref{znsystN2}) numerically using Mathematica.

\begin{tabular}{l}
\textbf{(\ref{znsystN2}a) Initial conditions} $z_1(0)=\hat{z}^{(1)}_{1}+0.01$, $z_2(0)=\hat{z}^{(1)}_{2}+0.01$,\\
 $\dot{z}_1(0)=0.01$, $\dot{z}_2(0)=-0.01$. See Figures~\ref{F11}, \ref{F12}, \ref{F13}, \ref{F14}.\\
\textbf{(\ref{znsystN2}b) Initial conditions} $z_1(0)=\hat{z}^{(1)}_{1}+0.01+0.01~\mathbf{i}$, $z_2(0)=\hat{z}^{(1)}_{2}+0.01+0.01~\textbf{i}$, \\
$\dot{z}_1(0)=-0.01$, $\dot{z}_2(0)=0.01$. See Figures~\ref{F21}, \ref{F22}.\\
\textbf{(\ref{znsystN2}c) Initial conditions} $z_1(0)=\hat{z}^{(2)}_{1}+0.01+0.01~\mathbf{i}$, $z_2(0)=\hat{z}^{(2)}_{2}+0.01+0.01~\mathbf{i}$,\\
 $\dot{z}_1(0)=-0.01$, $\dot{z}_2(0)=0.01$. See Figures~\ref{F31}, \ref{F32}.\\
\textbf{(\ref{znsystN2}d) Initial conditions} $z_1(0)=\hat{z}^{(3)}_{1}+0.01+0.01~\mathbf{i}$, $z_2(0)=\hat{z}^{(3)}_{2}+0.01+0.01~\mathbf{i}$,\\
 $\dot{z}_1(0)=-0.005$, $\dot{z}_2(0)=0.005$. See Figures~\ref{F41} and~\ref{F42}.\\
\textbf{(\ref{znsystN2}e) Initial conditions} $z_1(0)=\hat{z}^{(4)}_{1}+0.01+0.01~\mathbf{i}$, $z_2(0)=\hat{z}^{(4)}_{2}+0.01+0.01~\mathbf{i}$,\\
 $\dot{z}_1(0)=-0.01$, $\dot{z}_2(0)=0.01$. See Figures~\ref{F41} and~\ref{F42}.
\end{tabular}

\vspace{4mm}
 
\textbf{Example 2.2}.  For $N=3$ and $\omega=1$, taking into account that $c_1=-z_1-z_2-z_3$, $c_2=z_1 z_2+z_1 z_3+ z_2 z_3$, and $c_3=-z_1 z_2 z_3$, we see that system~(\ref{NewGold}) reduces to
\begin{subequations}
\begin{eqnarray}
&&\ddot{z}_1=\frac{2 \dot{z}_1 \dot{z}_2}{z_1-z_2}
+\frac{2 \dot{z}_1 \dot{z}_3}{z_1-z_3} \notag \\
&&-\frac{1}{(z_1-z_2)(z_1-z_3)}\left[
 z_1^2 F_1(z_1,z_2,z_3)+ z_1 F_2(z_1,z_2,z_3)+F_3(z_1, z_2, z_3))
 \right] , \notag\\
 &&\ddot{z}_2=-\frac{2 \dot{z}_1 \dot{z}_2}{z_1-z_2}
+\frac{2 \dot{z}_2 \dot{z}_3}{z_2-z_3} \notag \\
&&+\frac{1}{(z_1-z_2)(z_2-z_3)}\left[
 z_2^2 F_1(z_1,z_2,z_3)+ z_2 F_2(z_1,z_2,z_3)+F_3(z_1, z_2, z_3))
 \right] , \notag\\
&& \ddot{z}_3=-\frac{2 \dot{z}_1 \dot{z}_3}{z_1-z_3}
-\frac{2 \dot{z}_2 \dot{z}_3}{z_2-z_3} \notag \\
&&-\frac{1}{(z_1-z_3)(z_2-z_3)}\left[
 z_3^2 F_1(z_1,z_2,z_3)+ z_3 F_2(z_1,z_2,z_3)+F_3(z_1, z_2, z_3))
 \right] , \notag\\
 \label{znsystN3}
\end{eqnarray}
where
\begin{eqnarray}
&&F_1(z_1,z_2,z_3)=z_1 + z_2 + z_3 - \frac{2}{(z_1 + z_2 + z_3 - z_1 z_2 z_3)^3}\notag\\
&& - \frac{2}{(z_1+z_2 + z_3 +z_1 z_2+z_1 z_3+ z_2 z_3 )^3},\notag\\
&&F_2(z_1,z_2,z_3)=-z_1 z_2 - z_1 z_3 - z_2 z_3 \notag\\
&&+ \frac{2}{(z_1+z_2 + z_3 +z_1 z_2+z_1 z_3+ z_2 z_3 )^3}
 + \frac{2}{( z_1 z_2+z_1 z_3+ z_2 z_3+ z_1 z_2 z_3 )^3},\notag\\
&&F_3(z_1,z_2,z_3)=z_1 z_2 z_3 \notag\\
&&+ \frac{2}{(z_1+z_2 + z_3 -z_1 z_2 z_3 )^3} - \frac{2}{( z_1 z_2+z_1 z_3+ z_2 z_3+ z_1 z_2 z_3 )^3}.
\end{eqnarray}
\label{znsystN3Both}
\end{subequations}

 We obtained equilibria of system~~(\ref{znsystN3Both}) as follows. First, we found the equilibria of system~(\ref{cmdotdot}) for $N=3$; they are given by $(0,\sqrt{\frac{3}{2}},-\sqrt{\frac{3}{2}})$ and $(0,\mathbf{i}\sqrt{\frac{3}{2}},-\mathbf{i}\sqrt{\frac{3}{2}})$, up to the permutations of the three coordinates. Second, we found the zeros of the monic polynomials whose coefficients are the equilibria of system~(\ref{cmdotdot}) for $N=3$. These zeros are equilibriumum solutions of system~(\ref{NewGold}) because of how this system is constructed. Therefore, system~(\ref{znsystN3Both}) has at least 12 equilibrium configurations $(\hat{z}^{(j)}_1, \hat{z}^{(j)}_2, \hat{z}^{(j)}_3)$, $j=1,2,\ldots,12$, up to the permutations of  $\hat{z}^{(j)}_1$, $\hat{z}^{(j)}_2$ and $\hat{z}^{(j)}_3$, whose approximate numerical values are given below:
 \small
\begin{eqnarray}
\begin{array}{l}
(\hat{z}^{(1)}_{1}, \hat{z}^{(1)}_{2},  \hat{z}^{(1)}_{3})=
		(0.720239 - 0.575751 ~\mathbf{i}, 0.720239 + 0.575751 ~\mathbf{i}, -1.44048),\\
(\hat{z}^{(2)}_{1}, \hat{z}^{(2)}_{2},  \hat{z}^{(2)}_{3})=
		(0.397225 + 1.07661 ~\mathbf{i}, -1.12106 - 0.854451 ~\mathbf{i}, 0.72384 - 0.222154 ~\mathbf{i}),\\
(\hat{z}^{(3)}_{1}, \hat{z}^{(3)}_{2},  \hat{z}^{(3)}_{3})=
		(0.397225 - 1.07661 ~\mathbf{i}, 0.72384 + 0.222154 ~\mathbf{i}, -1.12106 + 0.854451 ~\mathbf{i}),\\
(\hat{z}^{(4)}_{1}, \hat{z}^{(4)}_{2},  \hat{z}^{(4)}_{3})=
		(0.709, -0.3545 - 1.2656 ~\mathbf{i}, -0.3545 + 1.2656 ~\mathbf{i}),\\
(\hat{z}^{(5)}_{1}, \hat{z}^{(5)}_{2},  \hat{z}^{(5)}_{3})=
		(-0.781352, 1.00305 - 0.749241 ~\mathbf{i}, 1.00305 + 0.749241 ~\mathbf{i}), \\
(\hat{z}^{(6)}_{1}, \hat{z}^{(6)}_{2},  \hat{z}^{(6)}_{3})=
		(0, 0.612372 - 0.921816 ~\mathbf{i}, 0.612372 + 0.921816 ~\mathbf{i}), \\
(\hat{z}^{(7)}_{1}, \hat{z}^{(7)}_{2},  \hat{z}^{(7)}_{3})=
		(-0.82853 - 0.22063 ~\mathbf{i}, 0.82853 - 0.22063 ~\mathbf{i}, 1.666 ~\mathbf{i}), \\
(\hat{z}^{(8)}_{1}, \hat{z}^{(8)}_{2},  \hat{z}^{(8)}_{3})=
		(0, -0.673004 + 1.52228 ~\mathbf{i}, 0.673004 - 0.297537 ~\mathbf{i}), \\
(\hat{z}^{(9)}_{1}, \hat{z}^{(9)}_{2},  \hat{z}^{(9)}_{3})=
		(0.82853 + 0.22063 ~\mathbf{i}, - 1.666 ~\mathbf{i}, -0.82853 + 0.22063 ~\mathbf{i}), \\
(\hat{z}^{(10)}_{1}, \hat{z}^{(10)}_{2},  \hat{z}^{(10)}_{3})=
		(0, -0.673004 - 1.52228 ~\mathbf{i}, 0.673004 + 0.297537 ~\mathbf{i}), \\
(\hat{z}^{(11)}_{1}, \hat{z}^{(11)}_{2},  \hat{z}^{(11)}_{3})=
		(-1.00305 + 0.749241 ~\mathbf{i}, -1.00305 - 0.749241 ~\mathbf{i}, 0.781352), \\
(\hat{z}^{(12)}_{1}, \hat{z}^{(12)}_{2},  \hat{z}^{(12)}_{3})=
		(0, -1.87718, 0.652438).
\end{array}
\label{EquiSystznN3}
\end{eqnarray}
\normalsize
It is possible that the system of algebraic equations characterizing the equilibria of~(\ref{znsystN3Both}) has additional solutions besides those listed above. A direct attempt to solve this system of algebraic equations using Mathematica was unsuccessful, and we did not deem the matter surficiently relevant to justify further investigations.

\begin{figure}
\includegraphics[width=11cm]{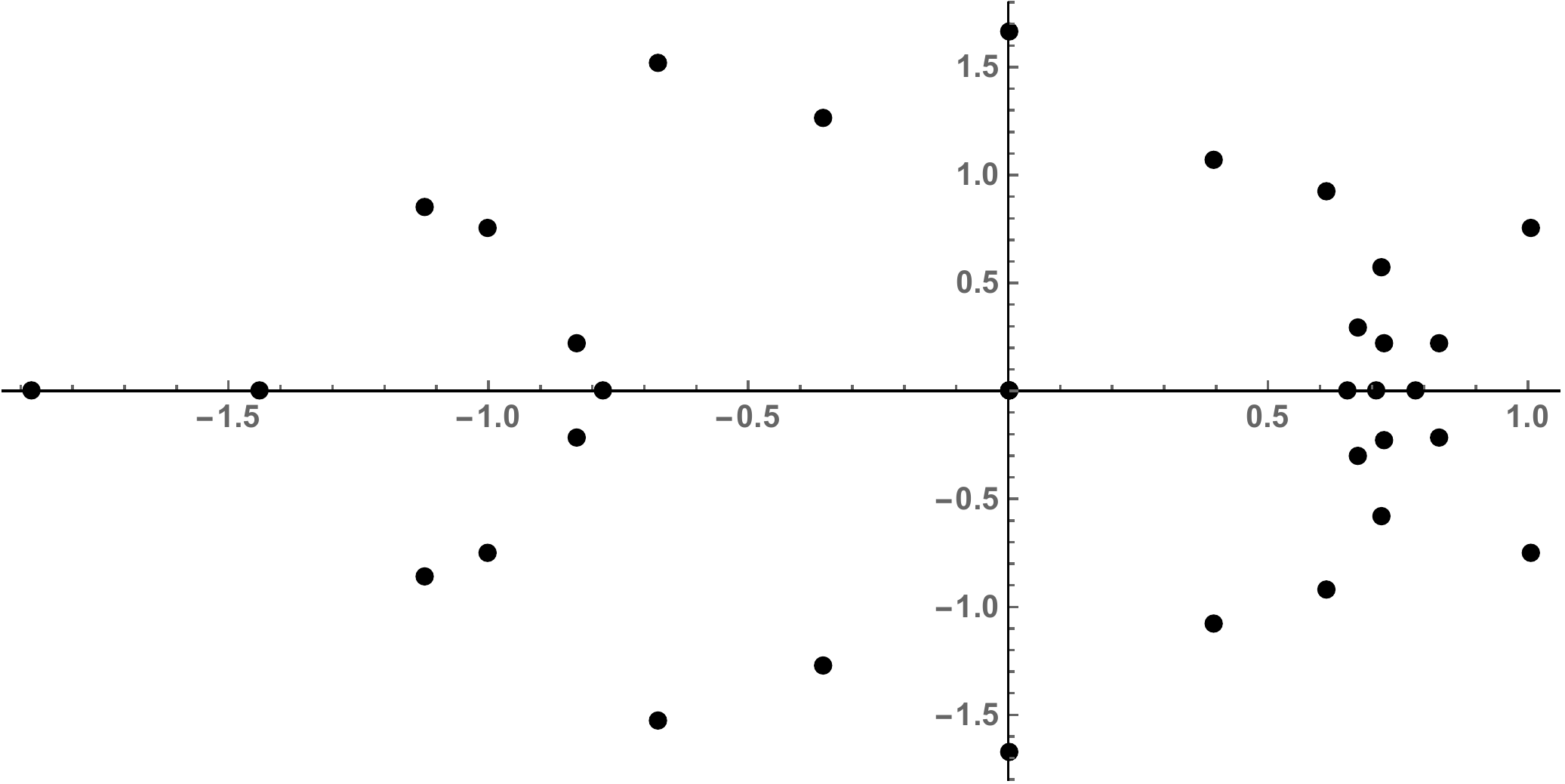}
\caption{Equilibria of system~(\ref{znsystN3Both}). Each equilibrium is represented by the three points $\hat{z}^{(j)}_{1}$, $\hat{z}^{(j)}_{2}$ and  $\hat{z}^{(j)}_{3}$ indexed by $j=1,2,\ldots,12$, see~(\ref{EquiSystznN3}).}
\label{EquilibriaN3}
\end{figure}

In Figure~\ref{EquilibriaN3} each equilibrium of system~(\ref{znsystN3Both}) is represented by the three points $\hat{z}^{(j)}_{1}$, $\hat{z}^{(j)}_{2}$ and $\hat{z}^{(j)}_{3}$, where $j=1,2,\ldots, 12$. We leave it to the interested reader to compare Figure~\ref{EquilibriaN3} with the list of equilibria~(\ref{EquiSystznN3}), in order to locate the triples that correspond to each of the 12 equilibria.

\begin{figure}
\includegraphics[width=11cm]{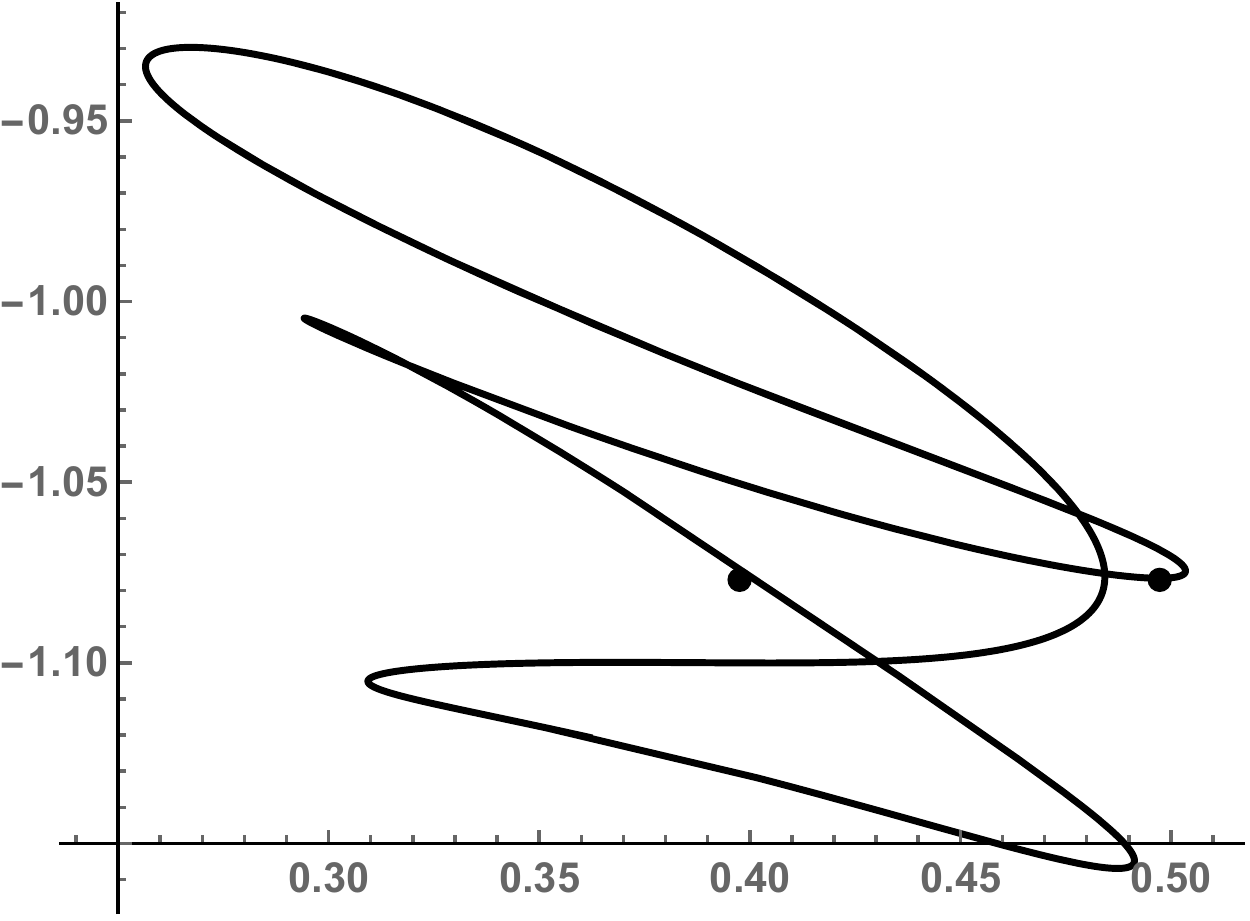}
\caption{
System~(\ref{znsystN3Both}), initial conditions~(\ref{InitCondN3}). Trajectory, in the complex $z$-plane, of $z_1(t)$. The two dots on the figure indicate the positions of  $\hat{z}^{(3)}_1$ respectively ${z}_1(0)$, i.e. of a nearby equilibrium point respectively the initial value of $z_1(t)$.}
\label{F61}
\end{figure}

\begin{figure}
\includegraphics[width=11cm]{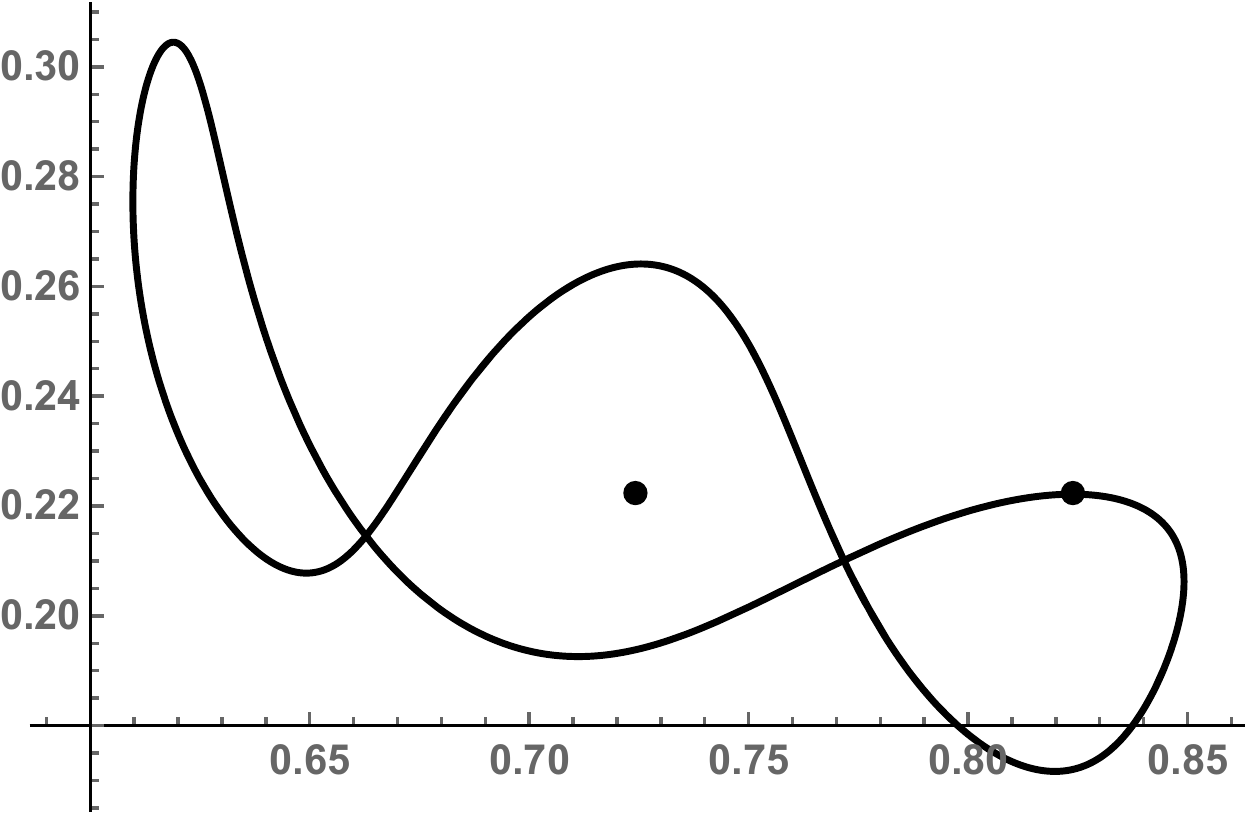}
\caption{
System~(\ref{znsystN3Both}), initial conditions~(\ref{InitCondN3}). Trajectory, in the complex $z$-plane, of $z_2(t)$. The two dots on the figure indicate the positions of  $\hat{z}^{(3)}_2$ and ${z}_2(0)$, i.e. of a nearby equilibrium point respectively the initial value of $z_2(t)$.}
\label{F62}
\end{figure}

\begin{figure}
\includegraphics[width=11cm]{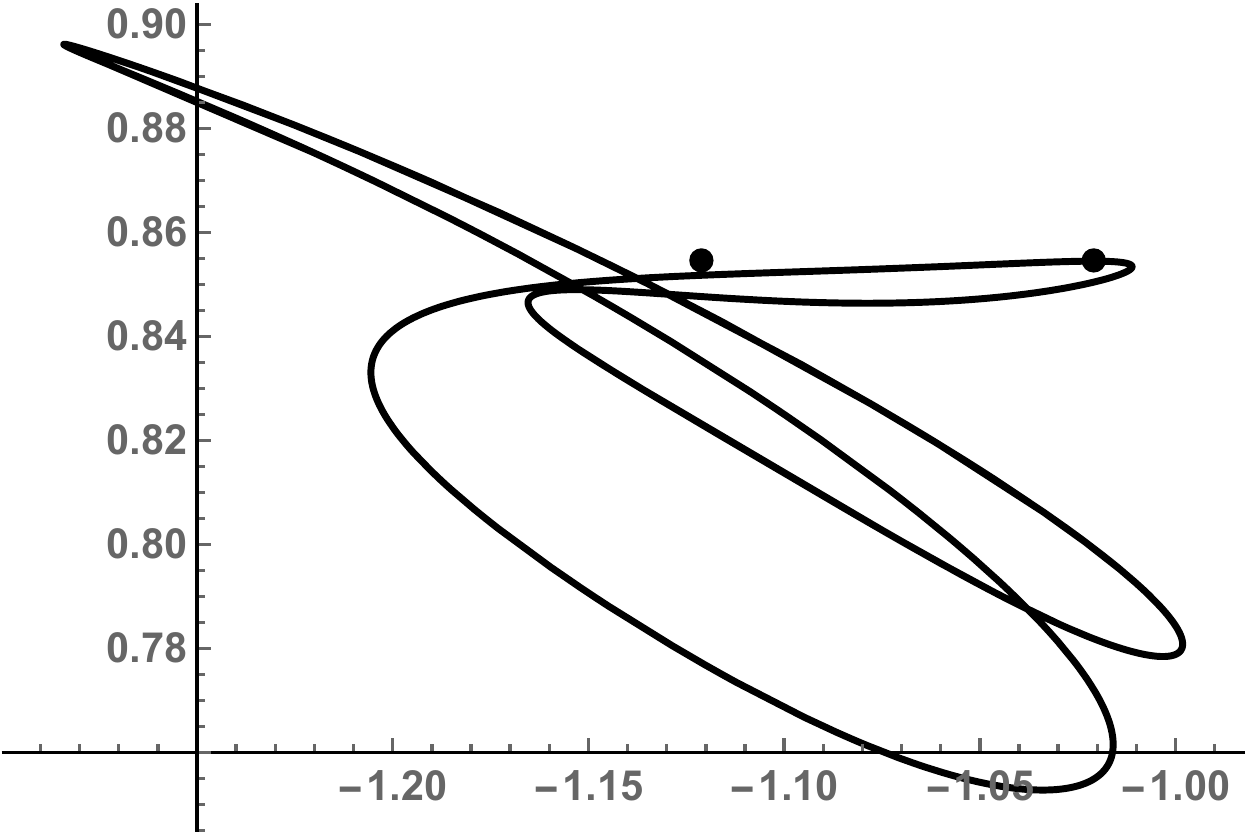}
\caption{
System~(\ref{znsystN3Both}), initial conditions~(\ref{InitCondN3}). Trajectory, in the complex $z$-plane, of $z_3(t)$. The two dots on the figure indicate the positions of  $\hat{z}^{(3)}_3$ and ${z}_3(0)$, i.e. of a nearby equilibrium point respectively the initial value of $z_3(t)$.}
\label{F63}
\end{figure}

Below we provide  
 some \textit{trajectories}, in the complex plane,
of the particles $z_1$, $z_2$ and $z_3$ whose evolution is described by system~(\ref{znsystN3Both}) with the initial conditions
\begin{eqnarray}
 z_1(0)=\hat{z}^{(3)}_1+0.1, \dot{z}_1(0)=0.1,\notag\\
 z_2(0)=\hat{z}^{(3)}_2+0.1, \dot{z}_2(0)=0.1,\notag\\
 z_3(0)=\hat{z}^{(3)}_3+0.1, \dot{z}_3(0)=0.1,
 \label{InitCondN3}
 \end{eqnarray} 
 see Figures~\ref{F61},~\ref{F62} and~\ref{F63}.
These graphs have been obtained by solving system~(\ref{znsystN3Both}) with the initial conditions (\ref{InitCondN3}) numerically using Mathematica.

\begin{figure}
\includegraphics[width=7cm]{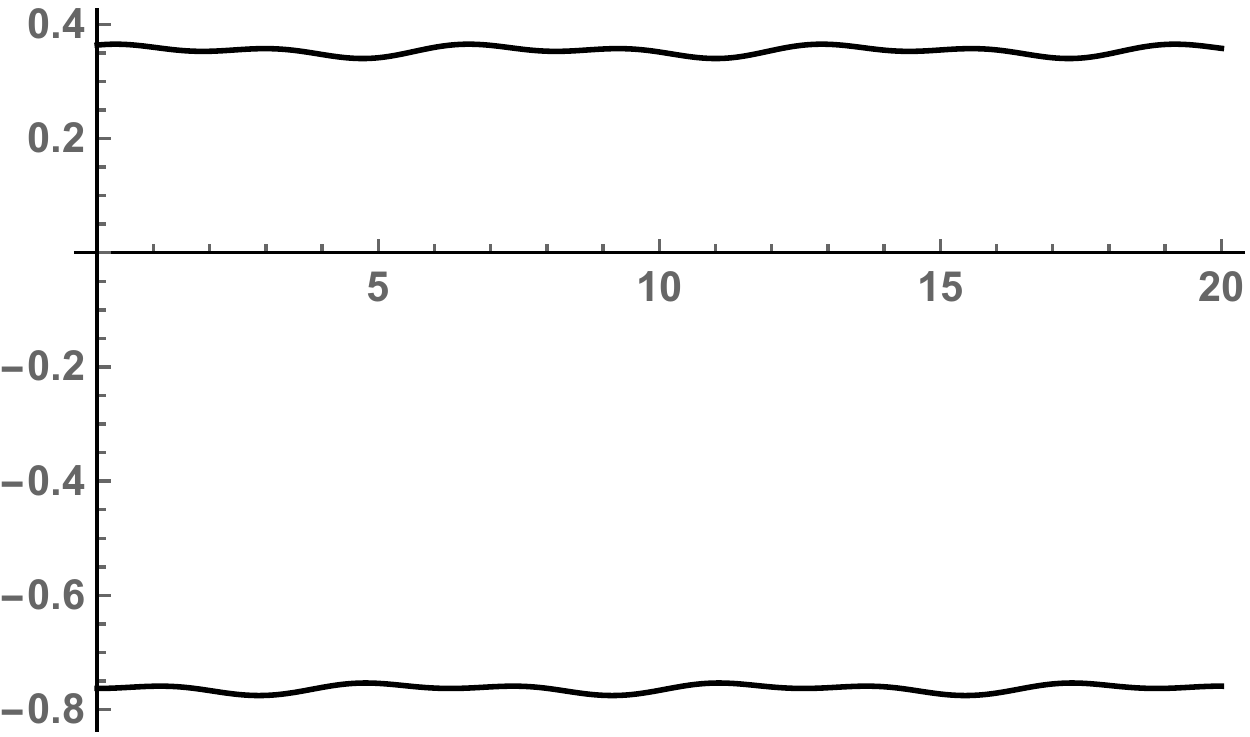}
\caption{ System~(\ref{znsystN2}), initial conditions~(\ref{znsystN2}a). Graphs of the real and imaginary parts of the coordinate $z_1(t)$.}
\label{F11}
\end{figure}

\begin{figure}
\includegraphics[width=7cm]{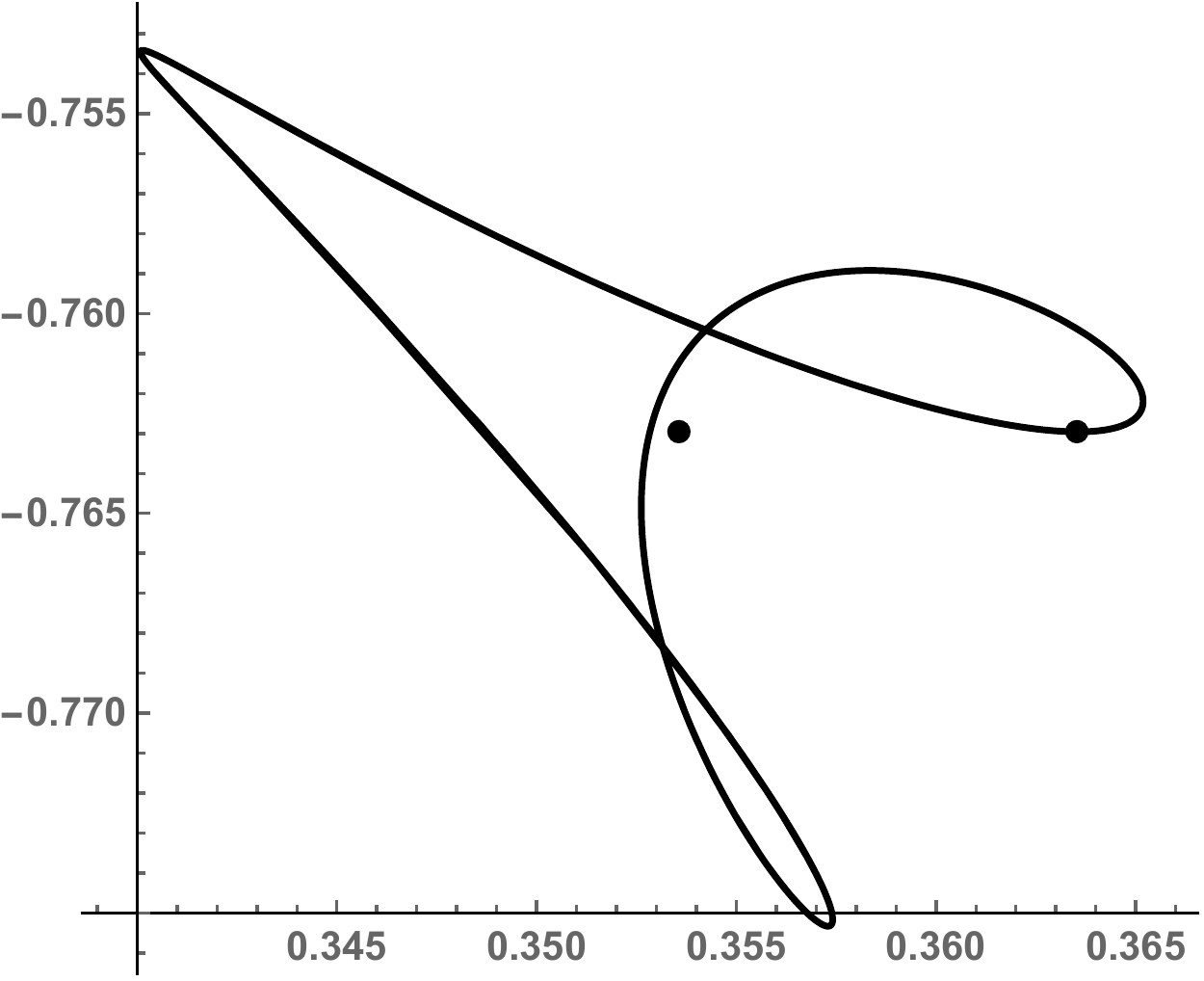}
\caption{System~(\ref{znsystN2}), initial conditions~(\ref{znsystN2}a). Trajectory, in the complex $z$-plane, of  $z_1(t)$. The two dots on the figure indicate the positions of  $\hat{z}^{(1)}_1$ respectively ${z}_1(0)$, i.e. of a nearby equilibrium point respectively the initial value of $z_1(t)$.}
\label{F12}
\end{figure}


\begin{figure}
\includegraphics[width=7cm]{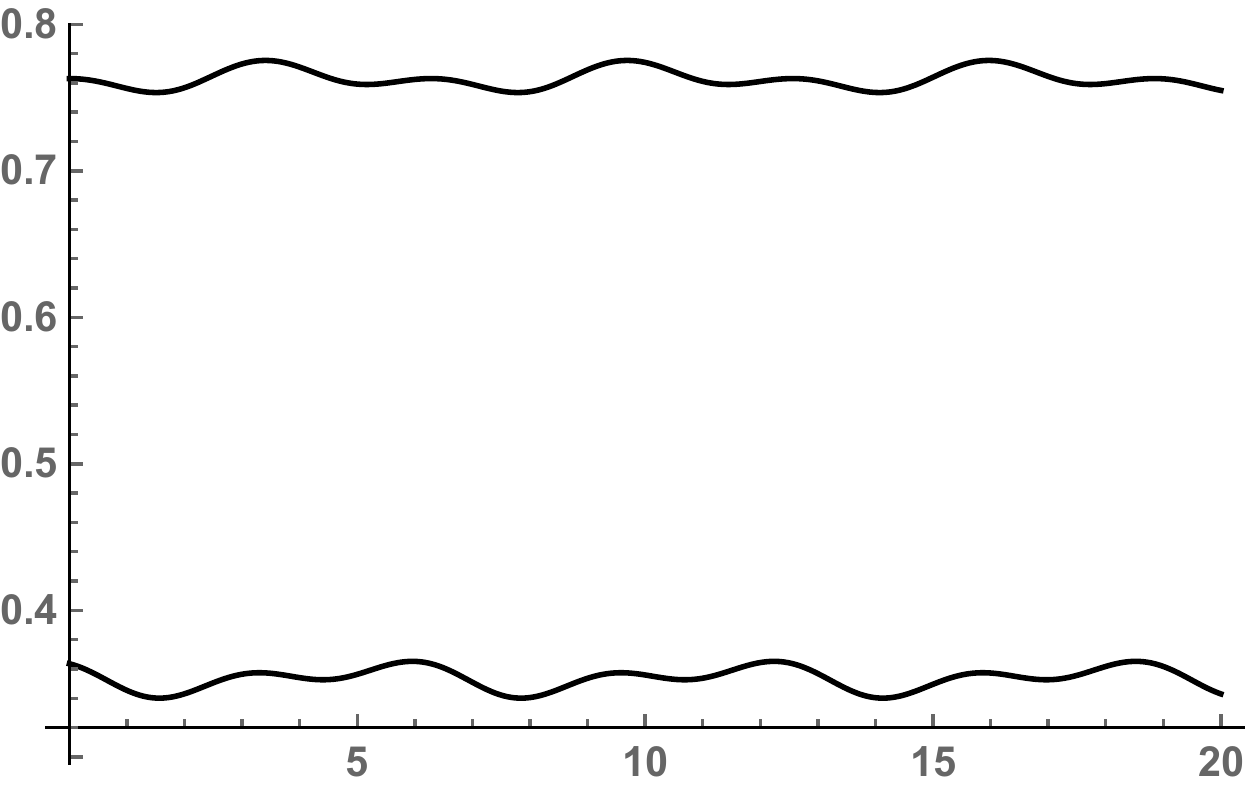}
\caption{
System~(\ref{znsystN2}), initial conditions~(\ref{znsystN2}a). Graphs of the real and the imaginary parts of the coordinate $z_2(t)$.}
\label{F13}
\end{figure}

\begin{figure}
\includegraphics[width=7cm]{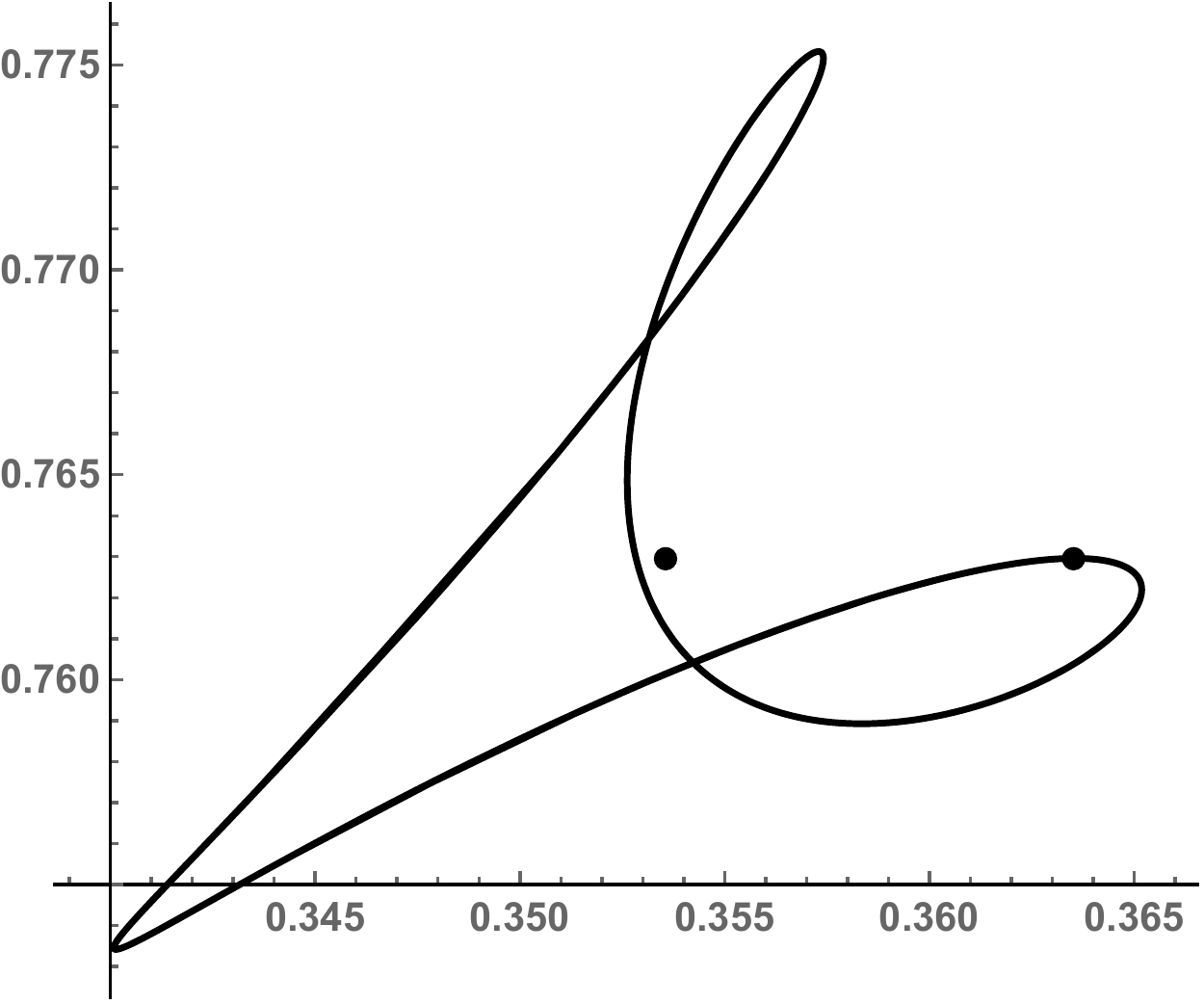}
\caption{
System~(\ref{znsystN2}), initial conditions~(\ref{znsystN2}a). Trajectory, in the complex $z$-plane, of $z_2(t)$. The two dots on the figure indicate the positions of  $\hat{z}^{(1)}_2$ respectively ${z}_2(0)$, i.e of a nearby equilibrium point respectively the initial value of $z_2(t)$.
}
\label{F14}
\end{figure}


\begin{figure}
\includegraphics[width=7cm]{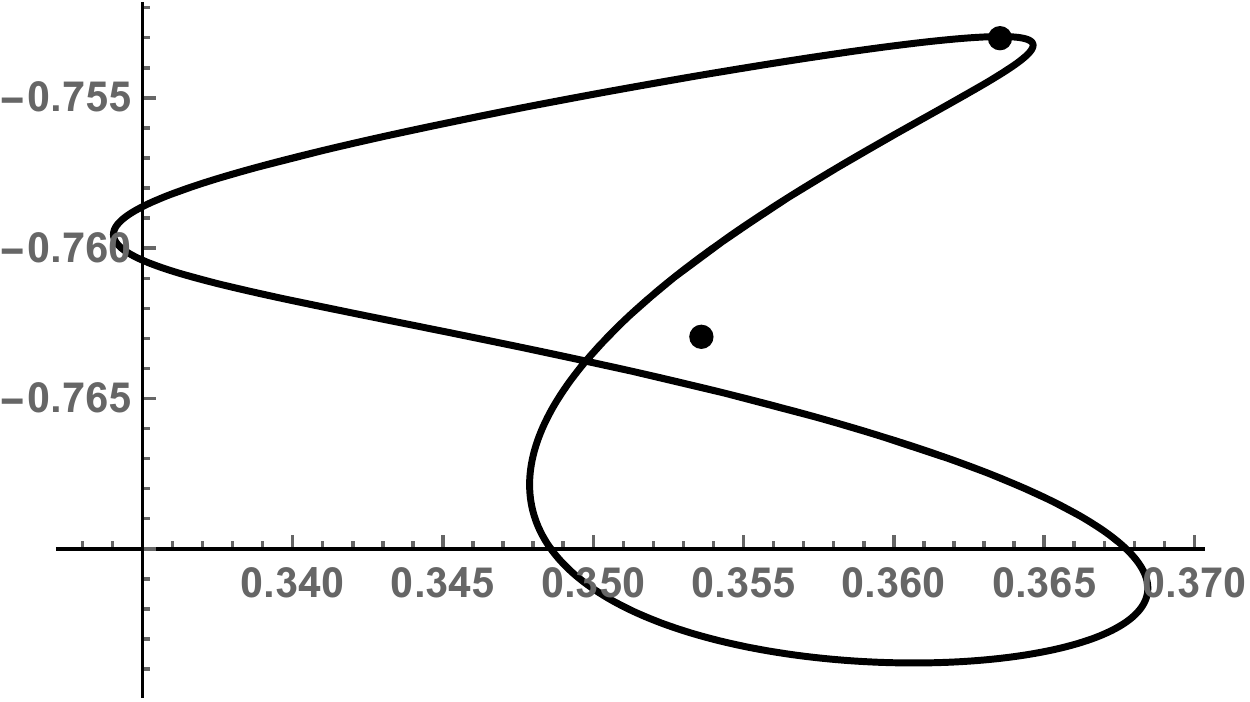}
\caption{
System~(\ref{znsystN2}), initial conditions~(\ref{znsystN2}b). Trajectory, in the complex $z$-plane, of  $z_1(t)$. The two dots on the figure indicate the positions of  $\hat{z}^{(1)}_1$ respectively ${z}_1(0)$,  i.e.  of a nearby equilibrium point respectively the initial value of $z_1(t)$.}
\label{F21}
\end{figure}

\begin{figure}
\includegraphics[width=7cm]{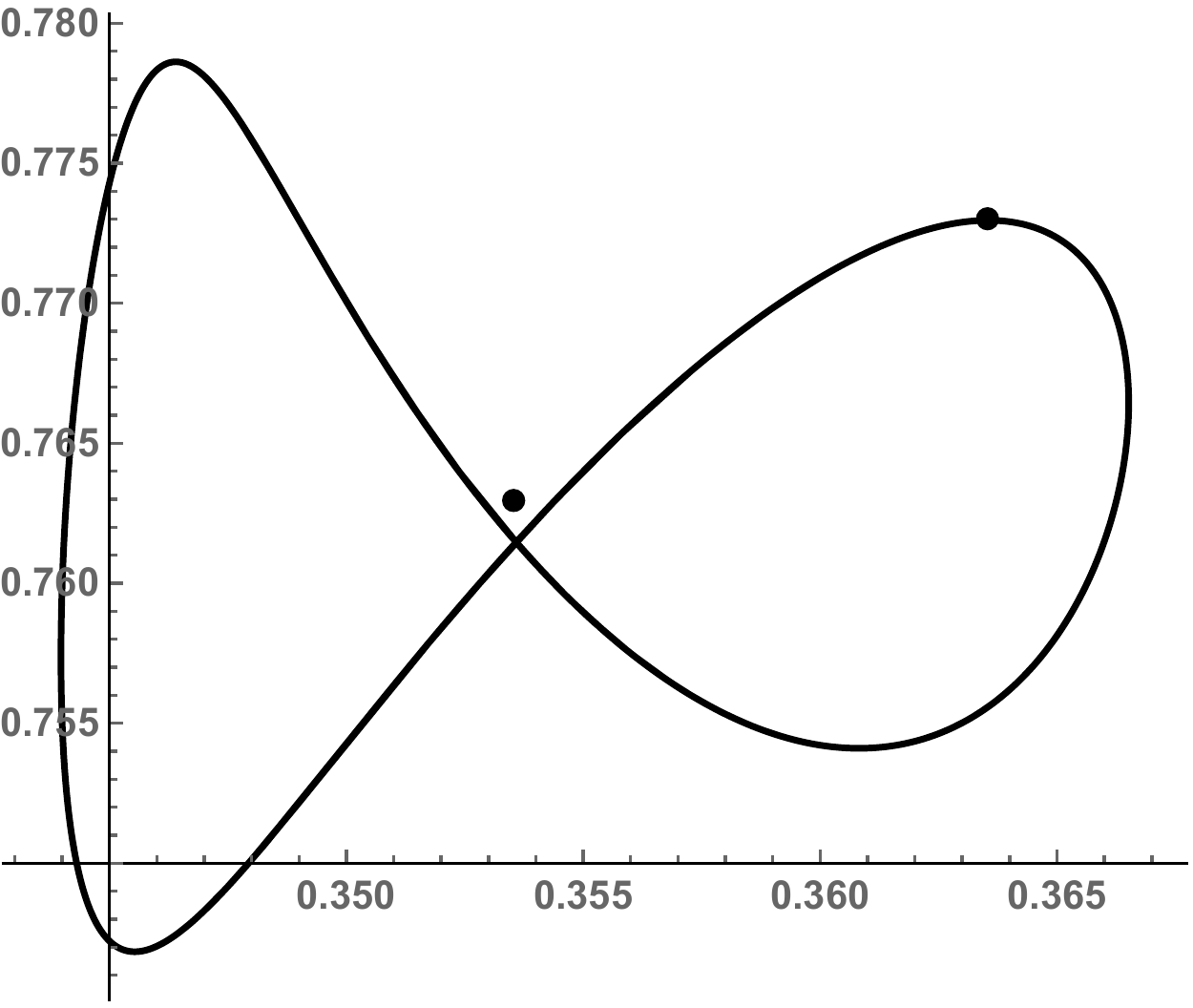}
\caption{
System~(\ref{znsystN2}), initial conditions~(\ref{znsystN2}b). Trajectory, in the complex $z$-plane, of $z_2(t)$. The two dots on the figure indicate the positions of  $\hat{z}^{(1)}_2$ respectively ${z}_2(0)$, i.e. of a nearby equilibrium point respectively the initial value of $z_2(t)$.}
\label{F22}
\end{figure}


\begin{figure}
\includegraphics[width=7cm]{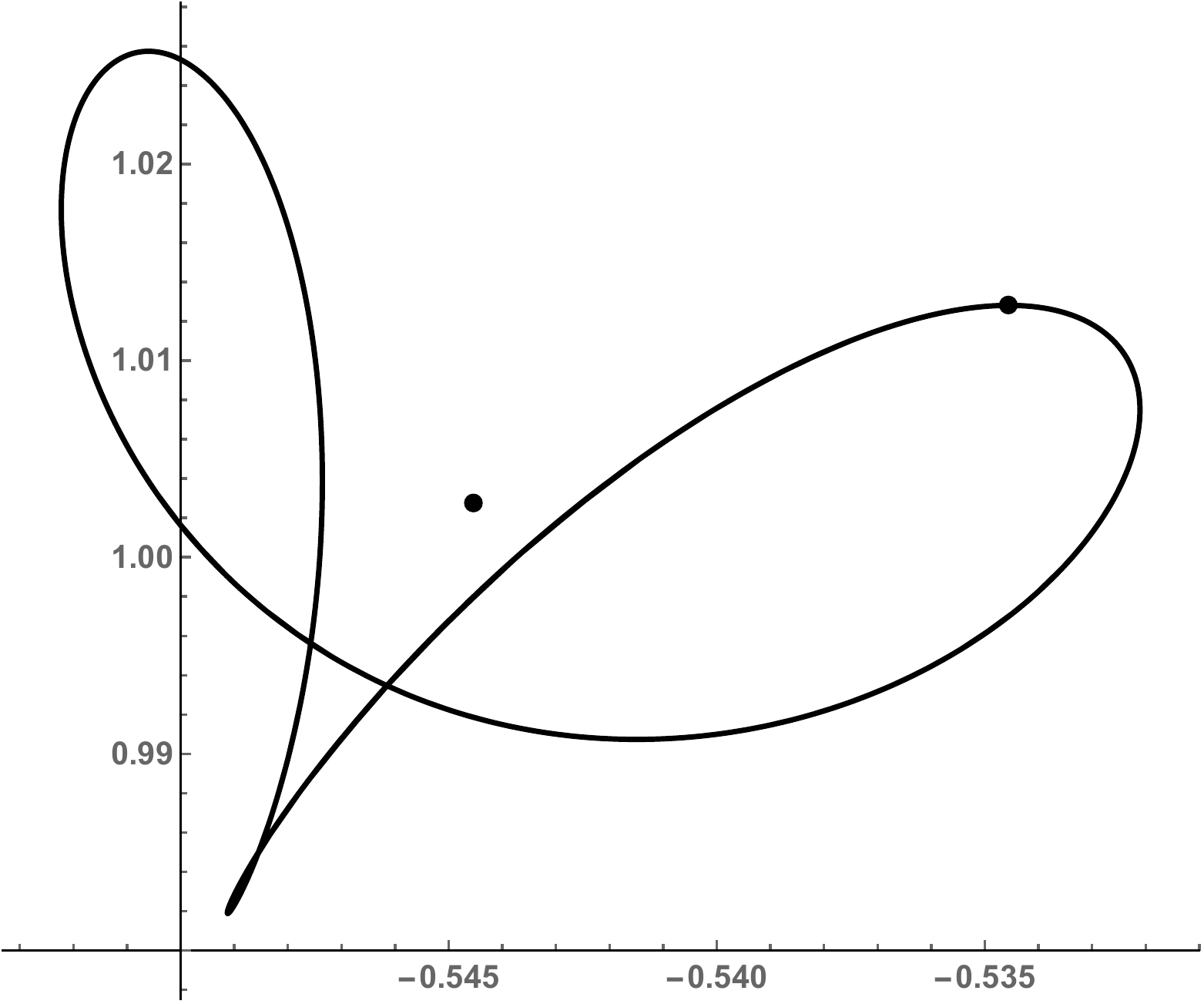}
\caption{
System~(\ref{znsystN2}), initial conditions~(\ref{znsystN2}c). Trajectory, in the complex $z$-plane, of  $z_1(t)$. The two dots on the figure indicate the positions of  $\hat{z}^{(2)}_1$ respectively ${z}_1(0)$, i.e. of a nearby equilibrium point respectively the initial value of $z_1(t)$.}
\label{F31}
\end{figure}

\begin{figure}
\includegraphics[width=7cm]{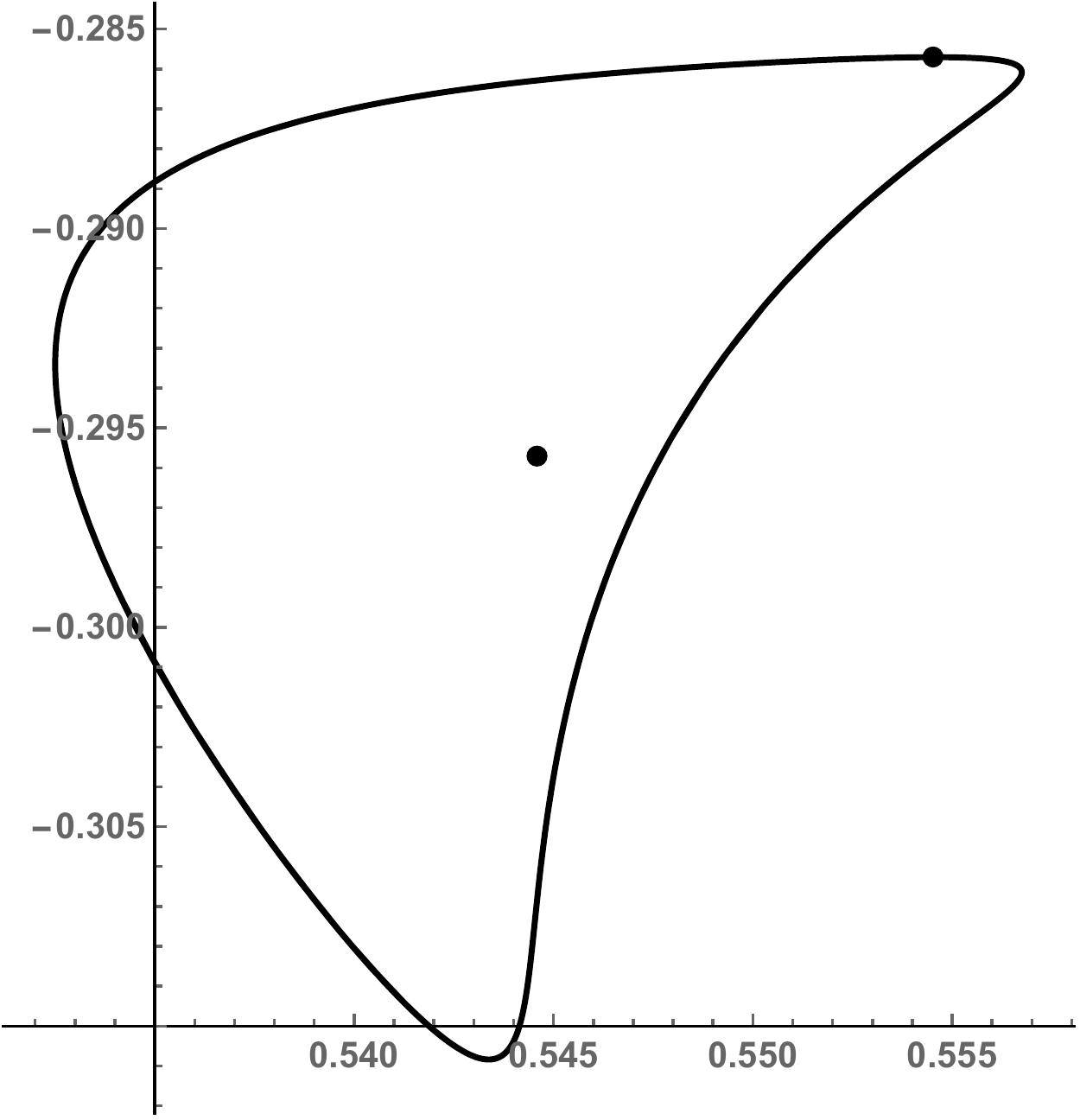}
\caption{
System~(\ref{znsystN2}), initial conditions~(\ref{znsystN2}c). Trajectory, in the complex $z$-plane, of $z_2(t)$. The two dots on the figure indicate the positions of  $\hat{z}^{(2)}_2$ respectively ${z}_2(0)$, i.e. of a nearby equilibrium point respectively the initial value of $z_2(t)$.}
\label{F32}
\end{figure}


\pagebreak 
\begin{figure}
\includegraphics[width=11cm]{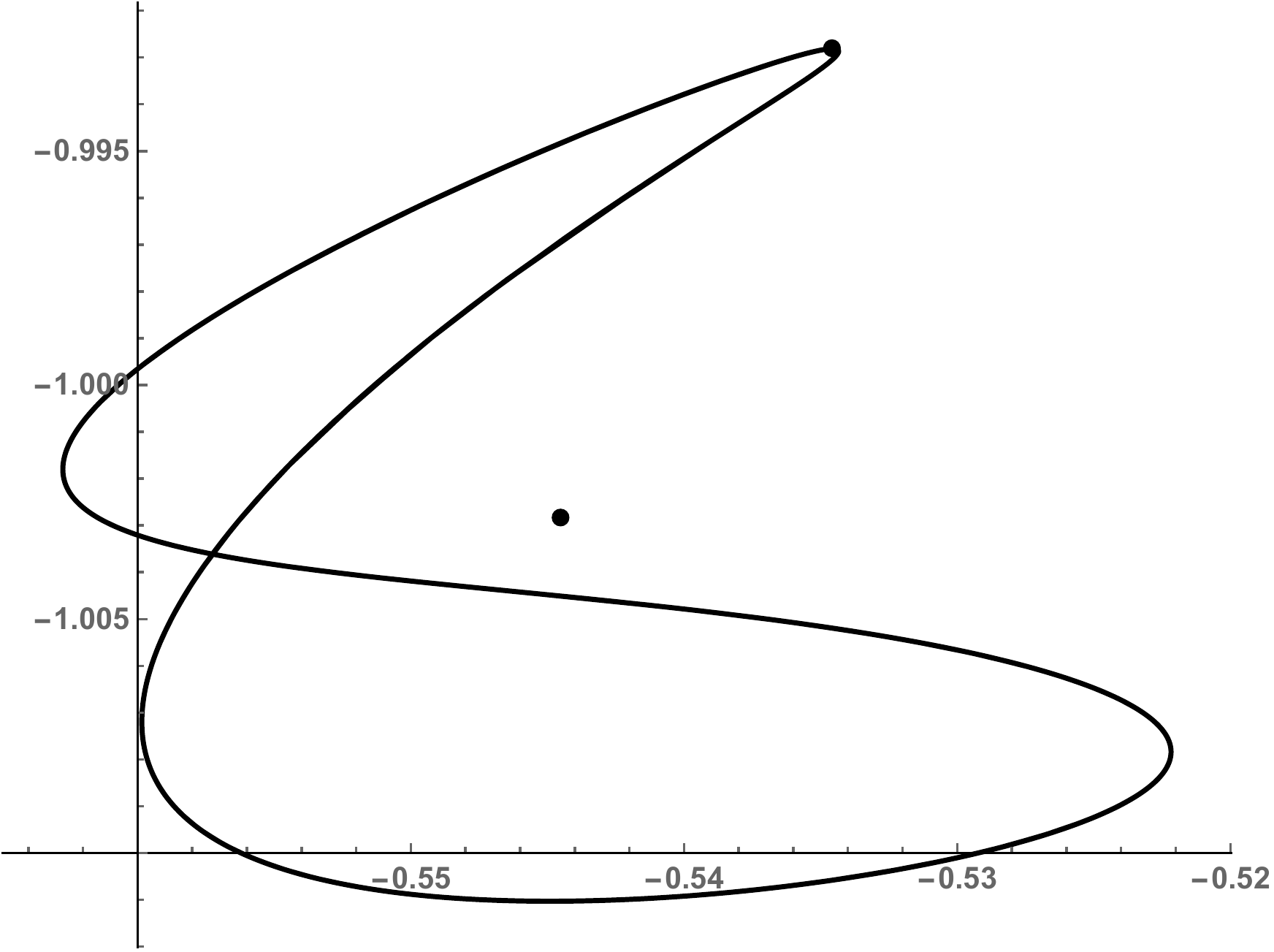}
\caption{
System~(\ref{znsystN2}), initial conditions~(\ref{znsystN2}d). Trajectory, in the complex $z$-plane, of $z_1(t)$. The two dots on the figure indicate the positions of  $\hat{z}^{(3)}_1$ and ${z}_1(0)$, i.e. of a nearby equilibrium point respectively the initial value of $z_1(t)$.}
\label{F41}
\end{figure}

\begin{figure}
\includegraphics[width=8cm]{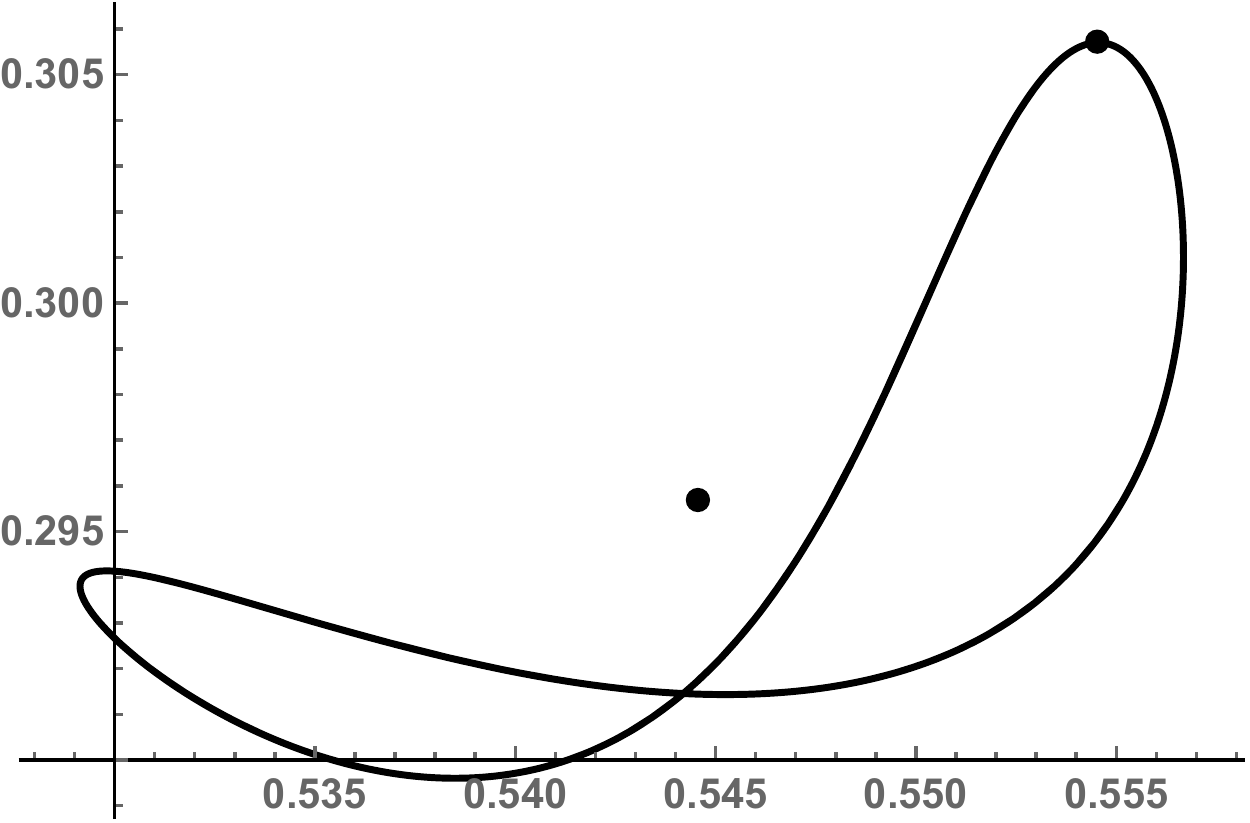}
\caption{
System~(\ref{znsystN2}), initial conditions~(\ref{znsystN2}d). Trajectory, in the complex $z$-plane, of  $z_2(t)$. The two dots on the figure indicate the positions of  $\hat{z}^{(3)}_2$ and ${z}_2(0)$, i.e. of a nearby equilibrium point respecttively the initial value of $z_2(t)$.}
\label{F42}
\end{figure}


\begin{figure}
\includegraphics[width=11cm]{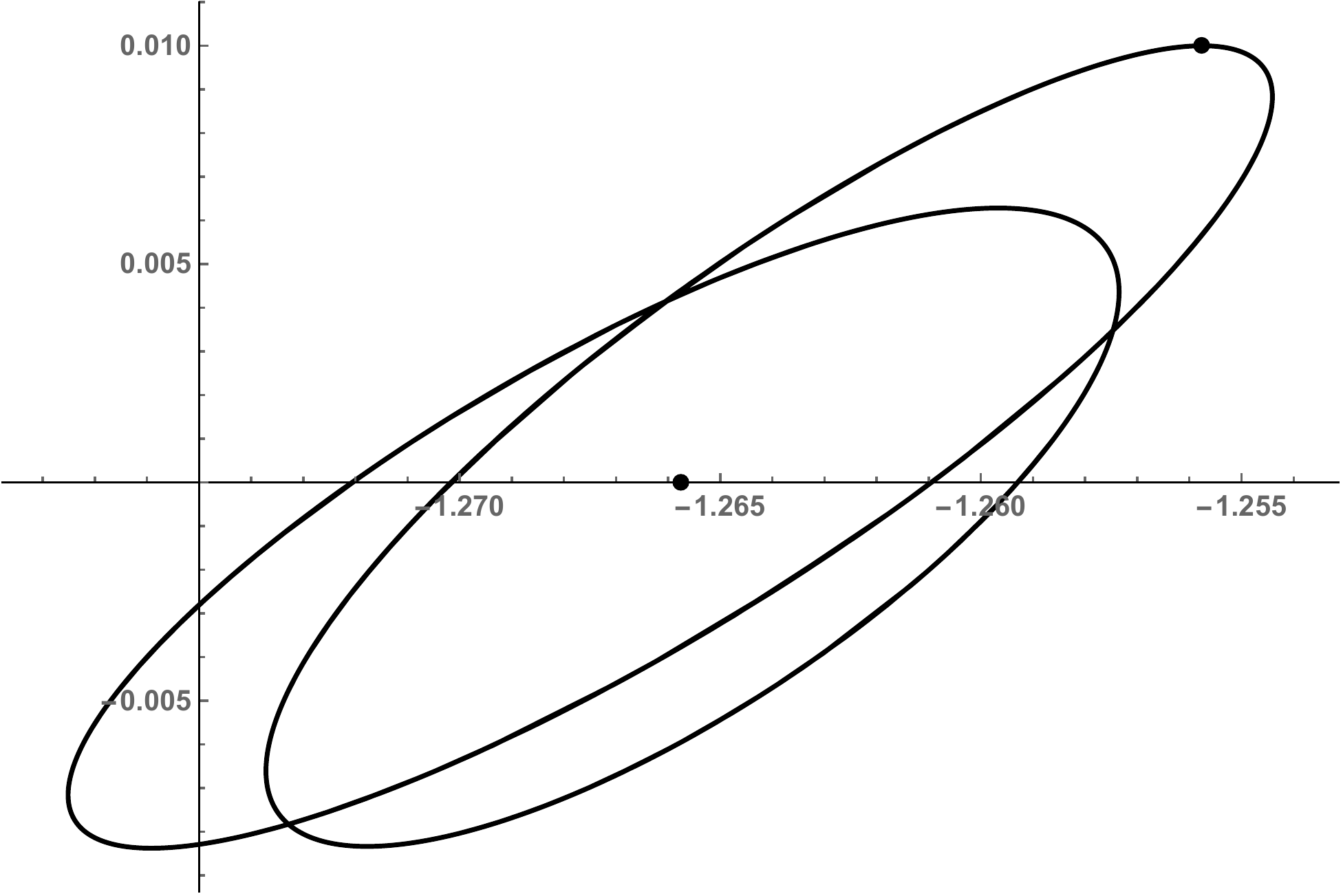}
\caption{
System~(\ref{znsystN2}), initial conditions~(\ref{znsystN2}e). Trajectory, in the complex $z$-plane, of $z_1(t)$. The two dots on the figure indicate the positions of  $\hat{z}^{(4)}_1$ and ${z}_1(0)$, i.e. of a nearby equilibrium point respectively the initial value of $z_1(t)$.}
\label{F51}
\end{figure}

\begin{figure}
\includegraphics[width=9cm]{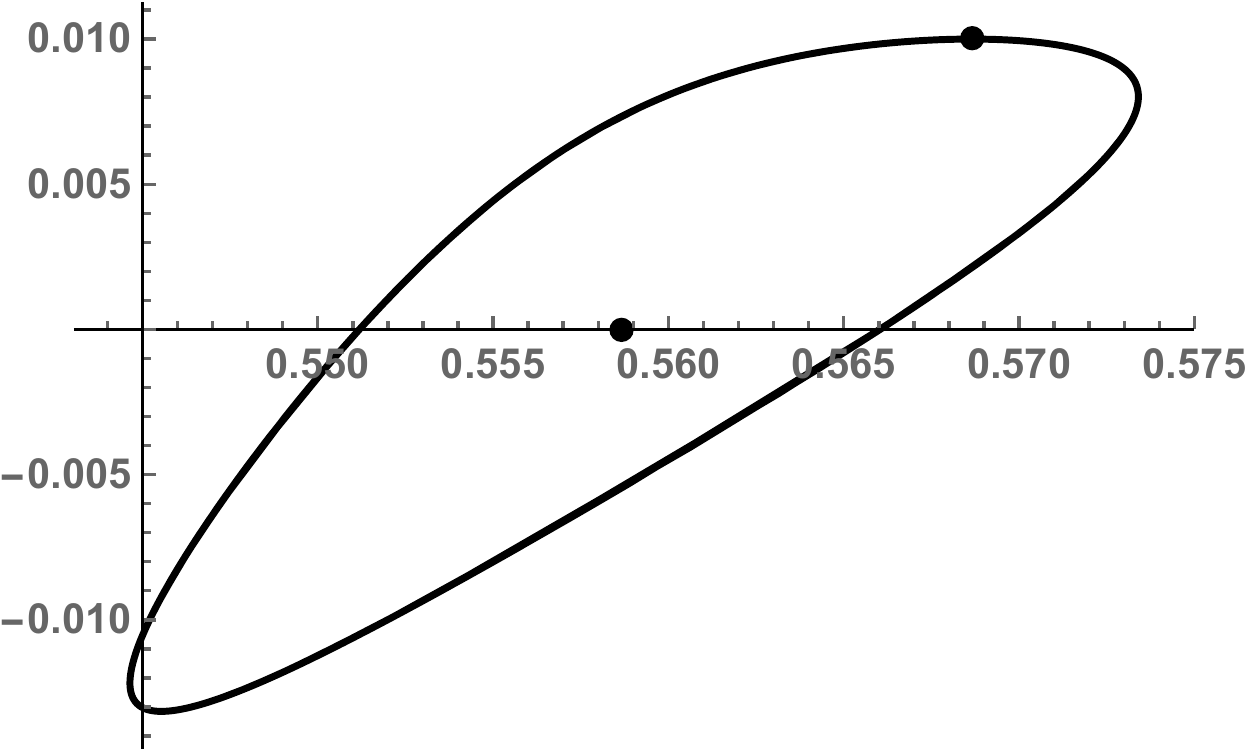}
\caption{
System~(\ref{znsystN2}), initial conditions~(\ref{znsystN2}e). Trajectory, in the complex $z$-plane, of $z_2(t)$. The two dots on the figure indicate the positions of  $\hat{z}^{(4)}_2$ and ${z}_2(0)$, i.e. of a nearby equilibrium point respectively the initial value of $z_2(t)$.}
\label{F52}
\end{figure}

\pagebreak

\section{Outlook}

The interest of the many-body model introduced and discussed
above is demonstrated by its \textit{solvable} and 
\textit{isochronous} character as well as by the remarkable trajectories 
it features -- already in the simple $N=2$ and $N=3$ cases, as shown by
the graphs reported above. It is also amusing to
observe---as the cognoscienti will have noted---that this solvable model has
been obtained by appropriately combining the two solvable equations---see (%
\ref{IsoGold}) and (\ref{cmdotdot})---which are the prototypes of two, quite
different, basic families of solvable many-body problems of Newtonian type.
Moreover---as reported in \cite{BC2015}---the findings reported above
provide the point of departure to obtain \textit{Diophantine} properties of
the $N$ zeros of each of the $N!$ (monic) polynomials the \textit{%
coefficients} of which are the $N$ \textit{zeros} of the Hermite polynomial $%
H_{N}\left( z\right) $ of degree $N$ (these $N!$ polynomials correspond of
course to the $N!$ permutations of the $N$ \textit{zeros} of $H_{N}\left(
z\right) $). And a rather ample vista of further developments is provided by
these findings.

\bigskip

\pagebreak

\end{document}